\documentclass[superscriptaddress,reprint]{revtex4-1} 
\usepackage{graphicx}

\usepackage{amsmath}    
\usepackage{color}
\usepackage{caption}

\usepackage{array}
\usepackage{booktabs}
\usepackage{multirow}
\newcommand{\head}[1]{\textnormal{\textbf{#1}}}
\newcommand{\normal}[1]{\multicolumn{1}{l}{#1}}

\usepackage{booktabs}

\begin{document}

\title{The role of water and steric constraints in the kinetics of cavity-ligand unbinding}

\author{Pratyush Tiwary} 
 \affiliation{Department of Chemistry, Columbia University, New York 10027, USA.}

\author{Jagannath Mondal} 
 \affiliation{Department of Chemistry, Columbia University, New York 10027, USA.}
\affiliation{TIFR-Centre for Interdisciplinary Sciences (TCIS), Tata Institute of Fundamental Research, Hyderabad 500075, India}

\author{Joseph A. Morrone} 
 \affiliation{Laufer Center for Physical and Quantitative Biology, Stony Brook University, Stony Brook, NY 11794}
 
\author{B. J. Berne} 
 \affiliation{Department of Chemistry, Columbia University, New York 10027, USA.}

%----------------------------------------------------------------------------------------

\begin{abstract}
A key factor influencing a drug's efficacy is its residence time in the binding pocket of the host protein. Using atomistic computer simulation to predict this residence time and the associated dissociation process is a desirable but extremely difficult task due to the long timescales involved. This gets further complicated by the presence of biophysical factors such as steric and solvation effects. In this work, we perform molecular dynamics (MD) simulations of the unbinding of a popular prototypical hydrophobic cavity-ligand system using a metadynamics based approach that allows direct assessment of kinetic pathways and parameters. When constrained to move in an axial manner, we find the unbinding time to be on the order of 4000 sec. In accordance with previous studies, we find that the ligand must pass through a region of sharp dewetting transition manifested by sudden and high fluctuations in solvent density in the cavity. When we remove the steric constraints on ligand, the unbinding happens predominantly by an alternate pathway, where the unbinding becomes 20 times faster, and the sharp dewetting transition instead becomes continuous.  We validate the unbinding timescales from metadynamics through a Poisson analysis, and by comparison through detailed balance to binding timescale estimates from unbiased MD. This work demonstrates that enhanced sampling can be used to perform explicit solvent molecular dynamics studies at timescales previously unattainable, obtaining direct and reliable pictures of the underlying physio-chemical factors including free energies and rate constants.
\end{abstract}

\maketitle % The \maketitle command is necessary to build the title page

\section{\large{Introduction}}

The unbinding of ligands from host substrates is a phenomenon widely occurring across biological and chemical sciences. It is of great interest to be able to understand the thermodynamics and kinetics of such processes, especially how they are influenced by solvent and steric effects. An accurate estimate of unbinding kinetics is in fact of crucial importance for drug discovery paradigms\cite{copeland,pan_kinetics}. However, in spite of the advent of massively parallel computer resources, it has not been so easy to simulate the dynamics of ligand unbinding and calculate associated rate constants.  The complications are mainly two fold. First, as has been seen in studies of model systems\cite{mondal_fuller,setny2013solvent,debenedetti_2015,morrone2012interplay,li2012hydrodynamic,sharma2012evaporation,giovambattista2006effect,mccammon_jacs},  various proteins\cite{young2007motifs,trypsin,hummer_review}, human immunodeficiency virus\cite{hiv} and actual anti-cancer drugs\cite{shaw_dasatanib,mondal_dasatanib,ladbury1996}, the solvent often manifests itself at the molecular scale. While coarse-grained models can be fit to explicit solvent molecular dynamics (MD) simulations\cite{mondal_fuller}, predictive power can be attained only by performing all-atom MD. Second, performing all-atom MD for unbinding of such systems is however plagued by the timescale problem. MD is restricted to integration timesteps of a few femtoseconds, which can be partially mitigated by multiple timestep MD algorithms\cite{tuckerman1992}. However, it is not yet routinely feasible to go into the millisecond regime and beyond for any system with more than few thousand atoms.

In this paper we consider a popular prototypical cavity-ligand system in explicit water where the attraction between water and the two nanoscale objects, namely a fullerene molecule and a spherical cavity, is weak \cite{mondal_fuller,setny2013solvent,morrone2012interplay,li2012hydrodynamic,sharma2012evaporation,giovambattista2006effect,mccammon_jacs}. We provide a full dynamical picture of the unbinding process demonstrating the clear role of water. We find that even in this relatively simple system there exists a rich range of dynamics that changes qualitatively and quantitatively as a function of the cavity-ligand distance and the motional degrees of freedom. Previous pioneering studies \cite{morrone2012interplay,li2012hydrodynamic,willard_tps,bolhuis_tps,sharma2012evaporation,mccammon_jacs,gardehummerprl} involving explicit all-atom MD, brownian dynamics, transition path sampling and other approaches have clearly shown that the association in such systems has a clear signature of solvent fluctuations and a sharp dewetting transition as the nanoscale objects approach each other -- if sterically constrained to move along the axis of symmetry. This is a popular set-up that has been considered in numerous studies over the years\cite{mondal_fuller,morrone2012interplay,li2012hydrodynamic,sharma2012evaporation,mccammon_jacs}, and is suggestive of biological systems where steric hindrances in the binding pocket do not allow the ligand to roll or move in a free manner\cite{trypsin,shaw_dasatanib}. Using unbiased MD and brownian dynamics tools, it was previously possible to calculate the timescales of association or binding for such systems that explicitly accounted for the dewetting transition\cite{mondal_fuller}.

 However, apart from one recent work \cite{trypsin}, to the best of our knowledge there is no reported study in which the timescales of the analogous dissociation or unbinding process were calculated through MD simulations. For such cavity-ligand systems due to the very high energy barrier of $\sim$30---40 $k_B T$\cite{mondal_fuller}, the unbinding timescales are simply too slow to be amenable through unbiased MD calculations. As such, in this work we use the popular enhanced sampling technique metadynamics\cite{meta_laio,wtm,tiwary_rewt,meta_ode} along with its recent extension\cite{meta_time,trypsin} for obtaining unbiased dynamics to calculate free energy profiles and unbinding rate constants for the cavity-ligand system. Furthermore, we ask and answer the following question: how does the dynamics of cavity-ligand association and dissociation depend on motional degrees of freedom? That is, how would the timescales of binding/unbinding vary between the cases where the ligand can/can not undergo free to move motion in any direction?

We find that the unbinding, in the case of motion being sterically constrained along the axis of symmetry of the system, proceeds through a sharp wetting transition at a critical ligand-cavity separation, conforming to the picture presented by Mondal et al through their position dependent friction calculation\cite{mondal_fuller}. The mean unbinding time of this system is found to be around 4000 seconds. The transition pathways harvested from our metadynamics-assisted MD runs are in perfect accordance with the previous calculations of Mondal et al \cite{mondal_fuller}. However, when the motion restraint is removed and the ligand is free to move in any direction, it finds an alternate pathway wherein there is no abrupt wetting transition, and the mean unbinding time reduces twenty-fold to 200 seconds. The binding times are also reduced. We validate rigorously all free energy profiles and rate constants through a combination of umbrella sampling, unbiased MD when feasible, and principle of detailed balance. The rate constants are further validated also through the Kolmogorov-Smirnov test for Poisson distribution proposed in Ref.\cite{pvalue}, thus giving further confidence in the estimated dynamics.

This work thus provides useful new insight into the phenomenon of hydrophobic interaction in solvated nanoscale systems\cite{berneweekszhou}, allowing one to directly simulate the unbinding process in MD in spite of the very high associated barriers. It also demonstrates that with a careful use of recently developed enhanced sampling techniques, one can perform molecular dynamics studies of unbinding/binding that extend well into the seconds timescale, and provide statistically accurate thermodynamic and kinetic information. The ideas that are used in this work are fairly generic and should be applicable to a large range of studies pertaining especially to ligand unbinding in explicit solvent.

%------------------------------------------------

\section{\large{Methods}}

 Throughout this work, we consider a cavity-ligand system (Fig. \ref{system}) where the attraction between water and both the nanoscale objects is weak and leaning towards a hydrophobic system (see SI for detailed potential forms and parameters). While the primary method in this paper is metadynamics, we directly and indirectly validate any findings from metadynamics with alternate independent approaches.
 
 \subsection{Metadynamics for free energy reconstruction}
\label{meta_fes}
Metadynamics is a widely used method for exploring complex free energy surfaces characterized by high free energy barriers\cite{meta_laio,wtm,meta_ode,tiwary_rewt,plumed1}. One first identifies a small number of slowly changing order parameters, called collective variables (CVs) \cite{meta_review}. A memory dependent biasing potential is constructed through the simulation as a function of these CVs, typically in the form of repulsive Gaussians added wherever the system visits in the CV space. Thus the system slowly starts to avoid the places where it has already visited. This leads to a gradual enhancement of the fluctuations in the CVs, through which the system is discouraged from getting trapped in the low free energy basins. At the end of a metadynamics run the probability distribution of any observable, whether biased directly or not, can be computed  through a reweighting procedure\cite{tiwary_rewt,bonomi_rewt}. This easy reweighting functionality is one of the many features of metadynamics that has made it a very popular method for calculation of free energy surfaces. 
 
 \begin{figure}[h!]
\includegraphics[width=1\columnwidth]{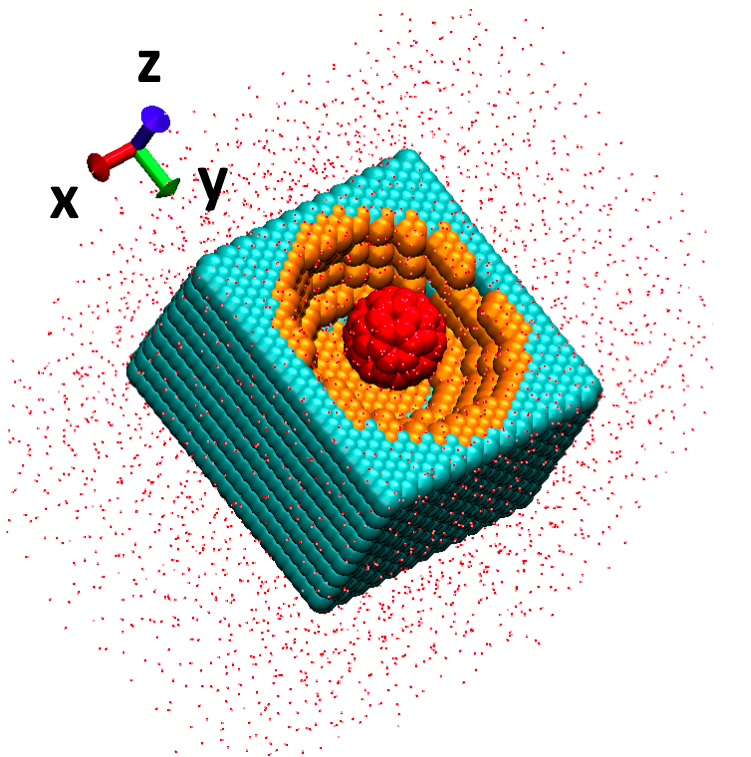}
\captionsetup{justification=raggedright,
singlelinecheck=false
}
\caption{Cavity-ligand system in explicit water with axes marked. Red: fullerene shaped ligand atoms. Orange: cavity atoms that interact with the ligand. Green: wall atoms. See SI for corresponding interaction potentials.}
\label{system}
\end{figure}

For the free energy reconstruction through metadynamics, in the case when the ligand is sterically constrained to move along the axis of symmetry ($z$-axis in Fig. \ref{system}), we perform one-dimensional metadynamics with the $z$-coordinate as the only CV. For the case when the ligand is free to move in any direction, we perform two-dimensional metadynamics with $z$- and the radial distance ($\rho = \sqrt{x^2 + y^2}$) from the axis of symmetry as the two CVs. In either case bias is added every 600 femtoseconds. In the SI we report the values of all other relevant parameters for both cases. In either case, we use restraining walls at high ligand-cavity separation to facilitate multiple re-entry events (see SI for details of restraints). 

\subsection{From metadynamics to dynamics}
\label{meta_kin}
Recently, Tiwary and co-workers extended the scope of metadynamics by showing how to extract unbiased rates from biased ones with minimal extra computational burden\cite{meta_time}. For this, they made two key assumptions on the dynamics:
\begin{enumerate}
\item  the transition processes are characterized by movements from one stable state to another via dynamical bottlenecks that are crossed only rarely, but when such a transition does happen, the time spent in the bottleneck is small
\item while there is no need to know beforehand the precise nature or location of these bottlenecks, one should have CVs that can distinguish between all stable basins of relevance. Note that this CV does not have to be the true reaction co-ordinate\cite{meta_time,besthummer_rc}.
\end{enumerate}

\begin{figure*}[t]
\label{fes_noroll}
  \centering
        \includegraphics[height=2.4in]{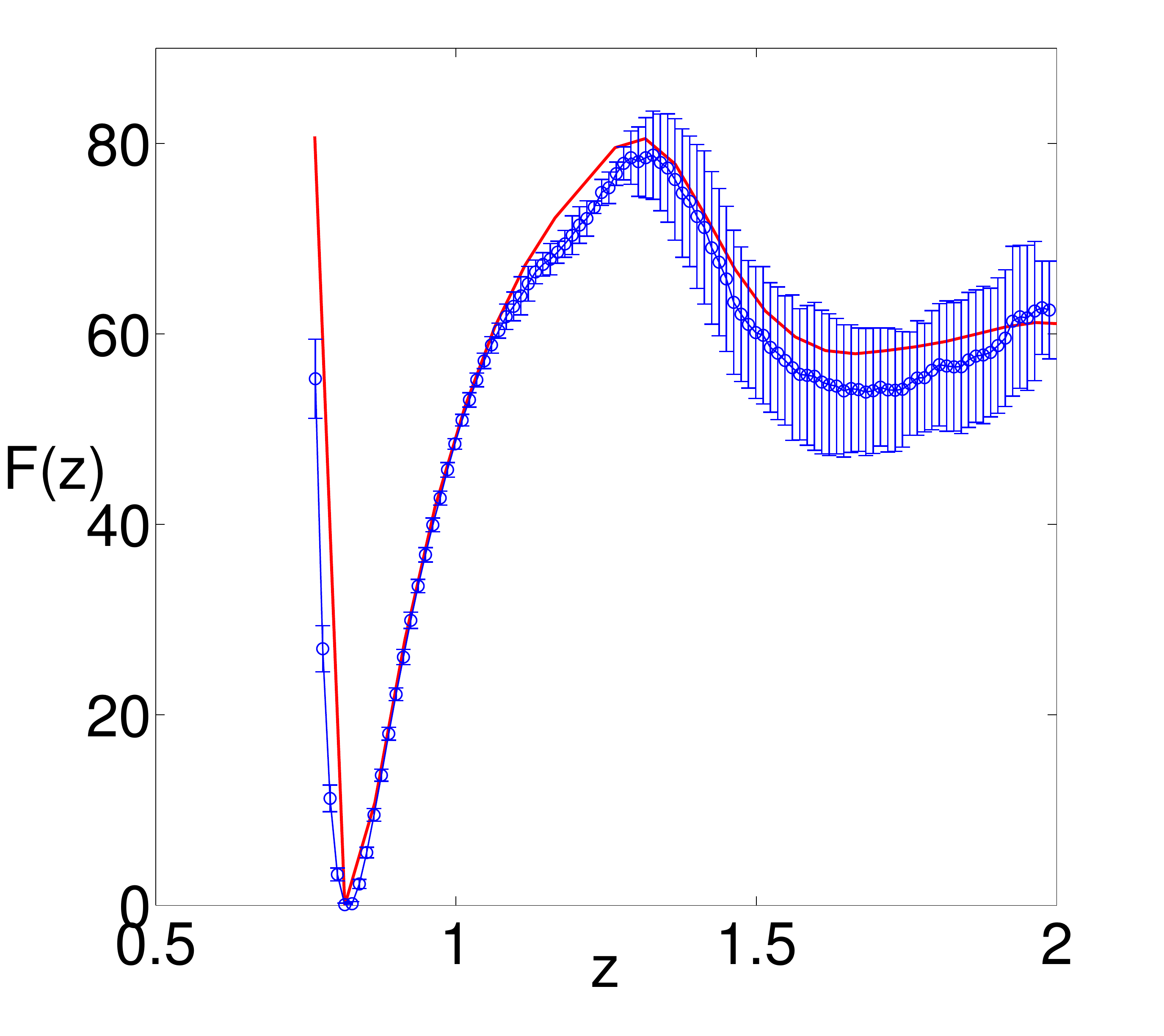} 
        ~
        \includegraphics[height=2.4in]{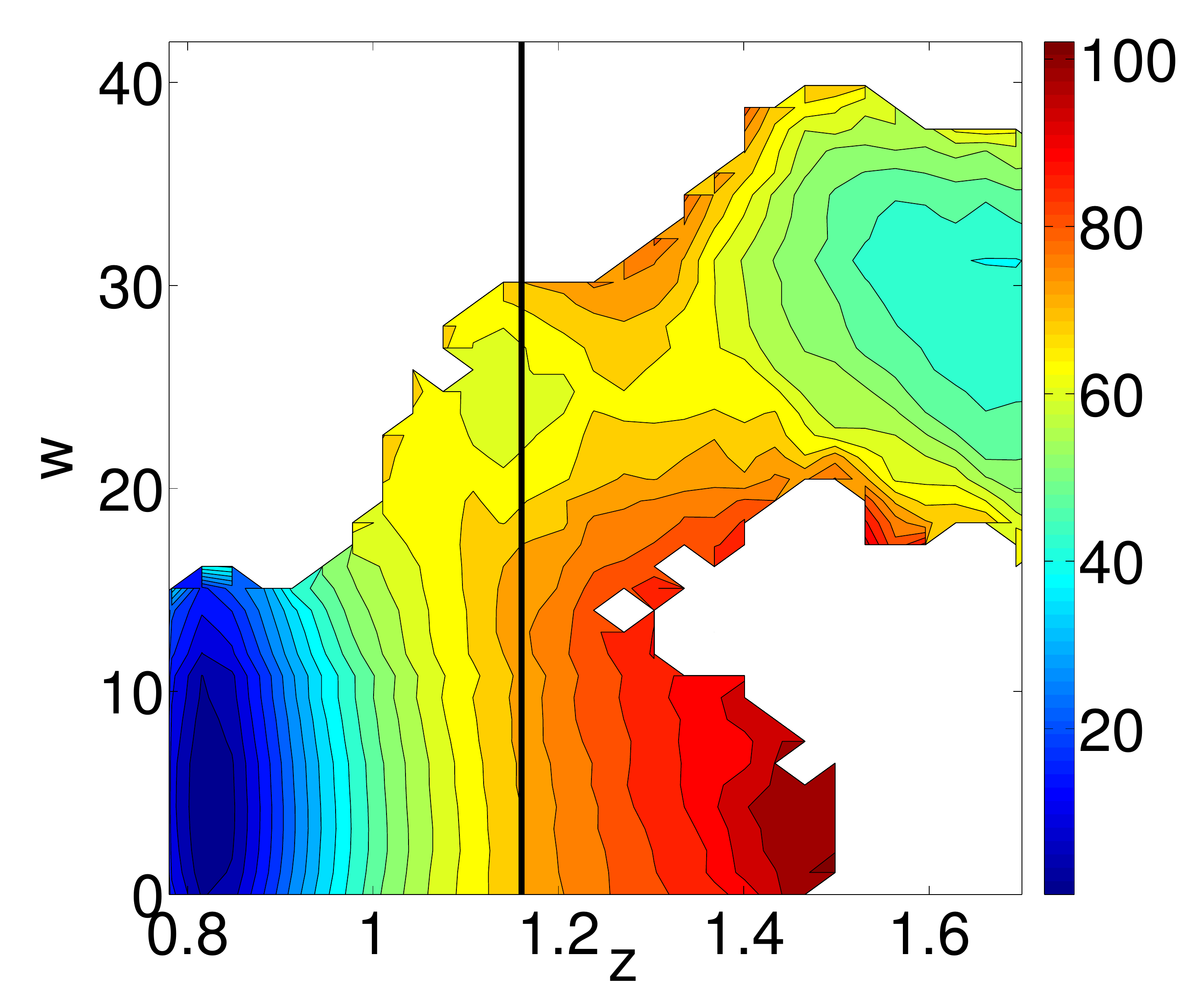}
        \captionsetup{justification=raggedright,
singlelinecheck=false
}
\caption{Free energy profiles when ligand is sterically constrained to move along system's axis of symmetry. (a) 1-d free energy as function of $z$. Blue denotes values from metadynamics with error bars and red from umbrella sampling. (b) 2-d free energy as function of $z$ and pocket water occupancy w (see \cite{mondal_fuller} for precise definition). The 2-d free energy was obtained by reweighting\cite{tiwary_rewt} 1-d metadynamics run performed along $z$. The solid black vertical line marks the critical $z_c$ value from Ref. \cite{mondal_fuller} at which dewetting transition had been predicted using a completely different approach. All energies are in kJ/mol. $z$ is in nm. }
\end{figure*}

Under these two key assumptions, by making the bias deposition slower than the time spent in dynamical bottlenecks, it is possible to keep the transition states (TS) relatively bias-free through the course of metadynamics. This so-called \textit{infrequent metadynamics} approach \cite{meta_time,pvalue,trypsin} preserves the unbiased sequence of state to state transitions and allows one to access the acceleration of transition rates achieved through biasing by appealing to generalized transition state theory\cite{berne1988} and calculating the following simple running average\cite{meta_time,voter_prl,grub_pre}:
\begin{equation}
\alpha = \langle e^{\beta V(s,t)} \rangle _t
\label{eq:acceleration}
\end{equation}
where $s$ is the collective variable being biased, $\beta$ is the inverse temperature, $V(s,t)$ is the bias experienced at time $t$ and the subscript $t$ indicates averaging under the time-dependent potential. The above expression is valid even if there are multiple intermediate states and numerous alternative reactive pathways\cite{meta_time,pvalue}.

In a successive work, a way was also proposed to assess the reliability of the two assumptions above\cite{pvalue}. This relies on the fact that the escape times from a long-lived metastable state obey a time-homogeneous Poisson statistics\cite{pvalue} with a single rate law. A statistical analysis based on the Kolmogorov-Smirnov (KS) test can quantitatively assess how precisely the above assumptions have been met \cite{pvalue}. Thus, if (a) significant bias got deposited in the TS region even with infrequent biasing, or (b) there are hidden unidentified timescales at play that the CV does not resolve, it would lead to failing the test for time-homogeneous Poisson statistics.

For the estimation of kinetics through metadynamics for the sterically constrained case, we perform one-dimensional metadynamics with the $z$-coordinate as the only CV (Fig. \ref{system}), but with a relatively much slower bias deposition rate of once every 10 ps. For the case when the ligand is free to move in any direction, we perform one-dimensional metadynamics with the overall ligand-cavity separation (i.e. $d=\sqrt {x^2 + y^2 + z^2}$) as CV and bias added every 10 ps. For each case, 14 independent simulations were started with the ligand fully bound to the cavity. The simulations were stopped when the ligand was fully unbound for the first time (see Results section for precise definition of unbound), and the respective unbinding times were calculated using the acceleration factor (Eq. \ref{eq:acceleration}). The transition time statistics so obtained was then subjected to a Poisson analysis to ascertain its reliability\cite{pvalue}. In the SI we report the values of all other relevant parameters for various cases.

One of the main features of infrequent metadynamics is that the segments of trajectories that cross the barrier between successive bias depositions are representatives of the unbiased transition path ensemble\cite{throwingropes}. As such we also present typical reactive trajectories for the sharp dewetting transition when it happens.

As a further verification of the unbinding timescale, we back-calculate the corresponding binding timescale using the free energy difference between the bound and unbound states and the principle of detailed balance. We compare this with the estimate of binding time from separate unbiased MD runs, given that binding is tractable through the latter. Specific details of this protocol are provided in the respective Results sections.

\subsection{ Umbrella sampling and related fixed bias method}

To further examine the quality of the free energies obtained from metadynamics, we perform independent umbrella-sampling based free energy calculations. For the scenario of ligand sterically constrained to move along the axis of symmetry, we use the method previously described by Mondal et al. \cite{mondal_fuller}. Briefly, in this method, one restrains the position of ligand along $z$ direction (Fig. \ref{system}) at a given ligand-pocket separation and computes the mean force being experienced by the ligand. By doing a  scan of mean force calculation along a wide-range of values of $z$ ($z$=0.8 to 2.0 nm), we obtain a profile of mean force acting between ligand and cavity along the $z$ direction. Finally, we integrate this mean-force-profile to obtain the  free energy profile along ligand-pocket separation $z$. The solvent-induced part of this free energy profile obtained in this fashion was already reported in our previous work\cite{mondal_fuller} and here we use the total free energy profile for comparison with metadynamics estimate.

For the scenario where there is no constraint on the movement of the ligand, we map the free energy using two-dimensional umbrella sampling. For this, we employ the same two CVs, namely the $z$- and $\rho=\sqrt {x^2 + y^2 }$ components of cavity-ligand distance, which we use in the corresponding metadynamics part as well (subsection ``Metadynamics for free energy reconstruction''). See SI and \cite{mondal_fuller} for further
details of the umbrella sampling protocol.

\section{\large{Results}}

We now systematically present the various results of our study. In Table 1, we summarize collectively the various timescales for binding and unbinding.
\label{constrained}
\subsection{Ligand sterically constrained to move along axis of symmetry}
\subsubsection{Thermodynamic profile}
In Fig. 2(a-b), we report the 1-dimensional free energy as a function of the distance $z$ between the centres of mass of the ligand and the substrate. The free energy profile as obtained by metadynamics is in very good agreement with that from umbrella sampling. We find an energy barrier in unbinding of around 80 kJ/mol. At an intermediate pocket-ligand separation, there is a relatively small but non-negligible barrier to binding as well, of around 20 kJ/mol. Our close inspection of the successful reactive trajectories (Fig. 3) shows that at this location there are large fluctuations in cavity-water density that must happen before the ligand unbinds from the host. Our previous prediction of dewetting-transition-mediated cavity-ligand recognition is confirmed by the  two-dimensional free energy surface as a function of the ligand-host separation $z$ and the number of water molecules in the binding pocket, as depicted in Fig. 2(b). Fig. 2(b) demonstrates that the sharp solvent fluctuation happens in a particular ligand-cavity separation of 1.15-1.25 nm, which is in very good agreement with previous estimate by Mondal et al \cite{mondal_fuller} using an independent approach.

\begin{table*}[!htbp]
\begin{tabular}{l*2{l}ll}
\bottomrule[1pt]
  & \multicolumn{2}{c}{\head{Unbinding (sec)}}
  & \multicolumn{2}{c}{\head{Binding (picosec)}}\\
  & \normal{\head{Sterically constrained}} & \normal{\head{free to move}}
  & \normal{\head{Sterically constrained}} & \head{free to move}\\
  \bottomrule[1pt]
  \multirow{1}{*}{Metadynamics:} &  3863 $\pm$ 1032  &   200 $\pm$ 51 & 769 $\pm$ 198 & 118 $\pm$ 31 \\
  \multirow{1}{*}{Unbiased MD:} &   N.A. &  N.A. & 476 $\pm$ 33 &   157 $\pm$ 5 \\
\bottomrule[1pt]
\end{tabular}
\captionsetup{justification=raggedright,
singlelinecheck=false
}
\caption{Binding and unbinding timescales in seconds and picoseconds respectively. The binding timescales from metadynamics denote values indirectly obtained after use of detailed balance with $\Delta G$ estimate from metadynamics.}
\end{table*}

\subsubsection{Kinetics}
Having established that 1-dimensional metadynamics performed using $z$ as a CV perfectly reproduces the salient thermodynamic features of this system including the  dewetting transition at the correct ligand-substrate separation, we then calculate the timescale of unbinding, which is the crux of the current article. For this, we use the infrequent metadynamics set-up, and calculate the acceleration factor achieved in metadynamics. 14 independent simulations were started with the ligand docked to the host, and stopped when $z=1.4$ was attained. We find a mean unbinding time of 3863  $\pm$ 1032 sec (Table 1). The transition time statistics so obtained fit very well a Poisson distribution \cite{pvalue} (see SI for detailed analysis), demonstrating that (1) the bias deposition was infrequent enough to not gradually corrupt the transition states, (2) biasing the CV $z$ was sufficient to ensure the time-scale separation needed for the infrequent metadynamics approach to be applicable. 

\begin{figure}[h!]
\label{reactraj}
  \centering
                \includegraphics[width=0.38\columnwidth]{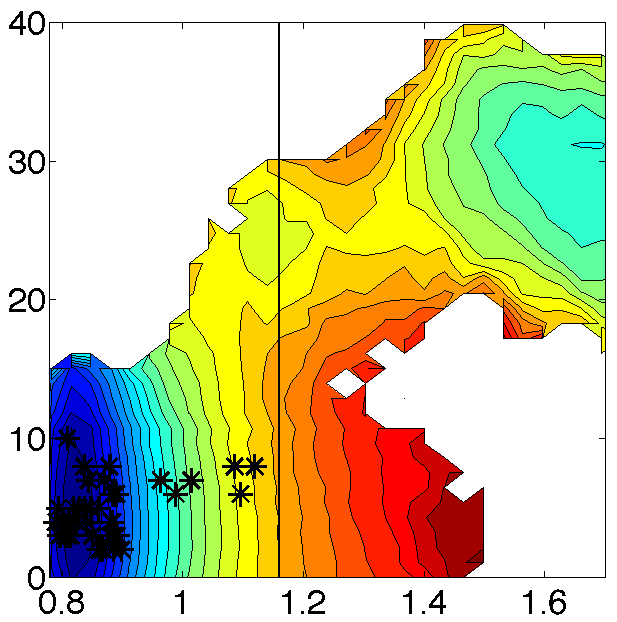}
        ~
                \includegraphics[width=0.38\columnwidth]{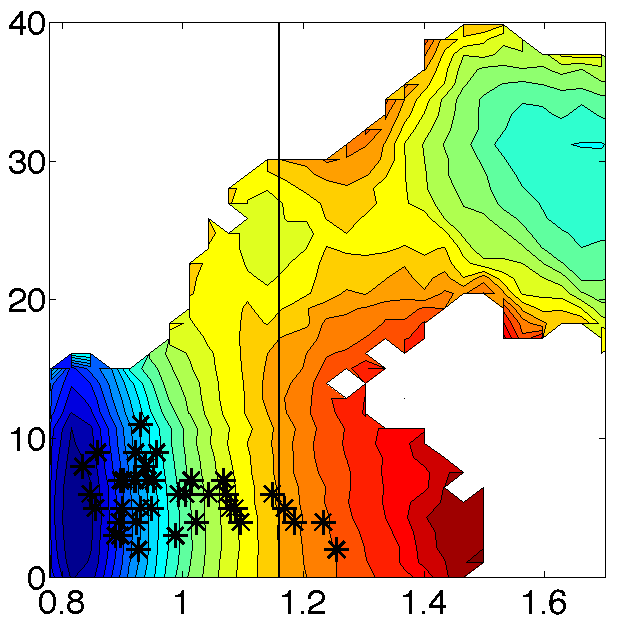}
                ~
                  \includegraphics[width=0.38\columnwidth]{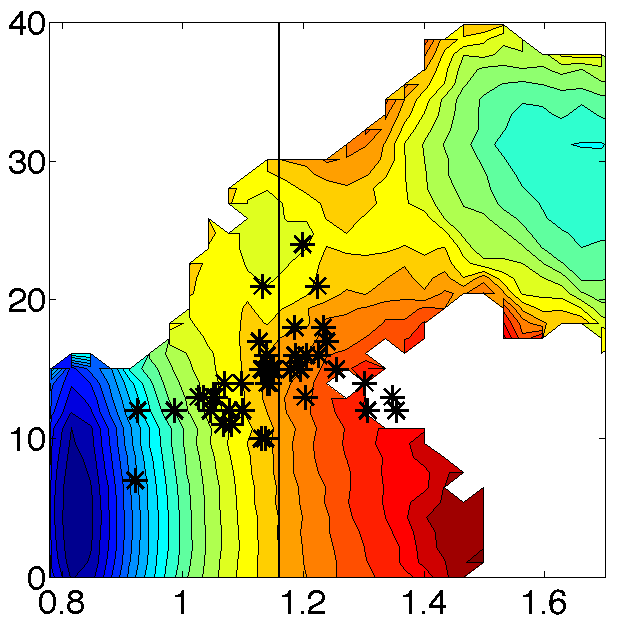}
                ~
                \includegraphics[width=0.38\columnwidth]{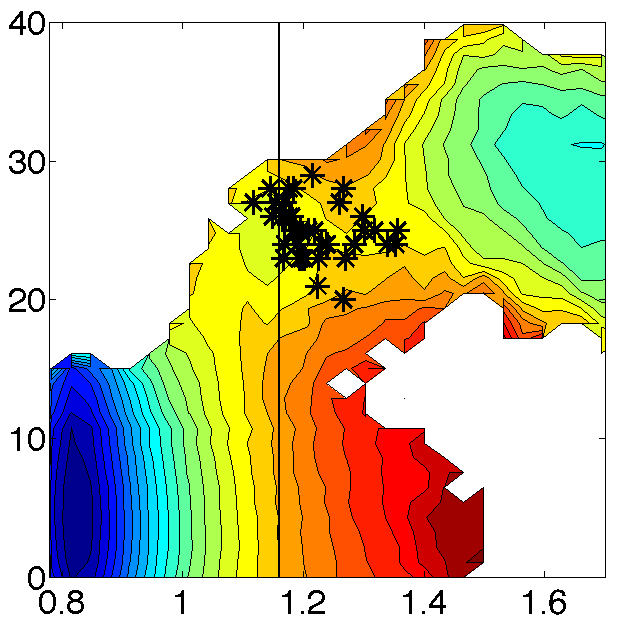}
                ~
                \includegraphics[width=0.38\columnwidth]{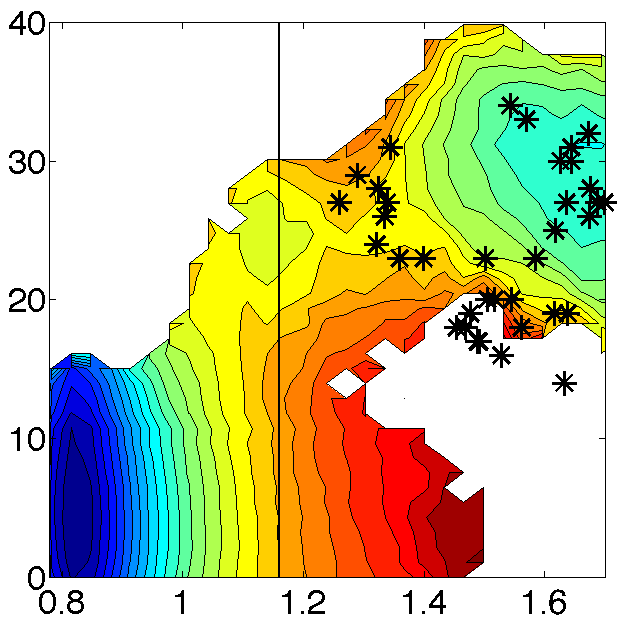}
                ~
                \includegraphics[width=0.38\columnwidth]{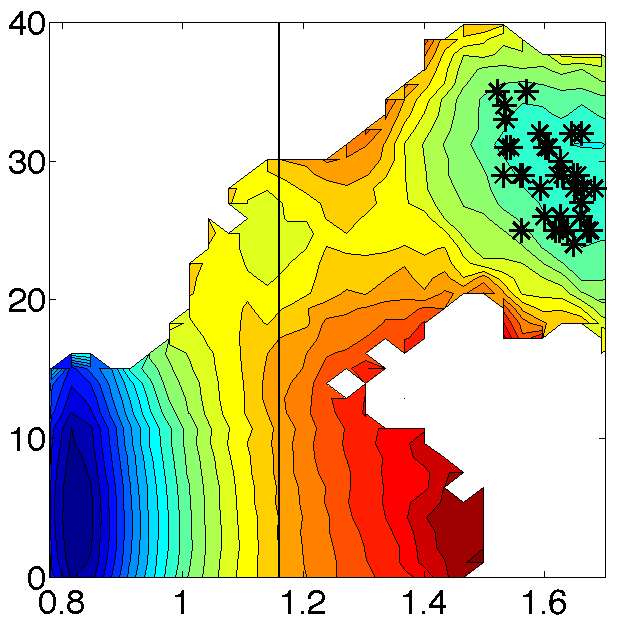}
                \captionsetup{justification=raggedright,
singlelinecheck=false
}
\caption{A typical reactive trajectory obtained via infrequent metadynamics providing a clear description of the dewetting transition as the sterically constrained ligand crosses over from bound to unbound state. Each snapshot proceeding row-by-row from top left to bottom right represents 10 ps long MD trajectory projected on ($z$,$w$) space. $x$-axis here is the $z$-distance, and $y$-axis is $w$, the number of pocket waters. The solid black vertical line marks the critical $z$-value at which Ref. \cite{mondal_fuller} had predicted the existence of large solvent fluctuations. The underlying free energy surface, axes and the contours are same as in Fig. 2(b).}
\end{figure}

To further validate the unbinding time estimate of 3863$\pm1032$ sec, we invoke the principle of detailed balance. We use the 1-d free energy profile (Fig. 2) and the relation  unbinding time = binding time $\times e^{-\beta \Delta G}$, where  $\Delta G$ is the free energy difference between the solvated and bound ligand. Here the mean binding time means the time for the fully solvated ligand ($z=1.4$) to get bound. Note that for the purpose of checking detailed balance, as is our objective here, any $z-$value would be fine. As per Fig. 2,  the estimate of $\Delta G$= -72.5 kJ/mol from metadynamics, giving us a  mean unbinding time of 769 $\pm$ 198 ps, which is well within order of magnitude agreement with the value of 476  $\pm$ 33 ps through unbiased MD simulations reported by Mondal et al \cite{mondal_fuller}.

In Fig. 3 we provide a set of snapshots of a typical $\sim$60ps long reactive trajectory as the system moves successfully from bound to unbound state. Here each snapshot corresponds to 10 ps of MD projected on the 2-dimensional (z,w) space, where w is the number of pocket waters (see \cite{mondal_fuller} for precise definition of w). This provides extremely clear dynamical evidence of the dewetting transition in such a hydrophobic and sterically constrained system. The solid black vertical line in  Fig. 3 marks the critical $z$-value at which Ref. \cite{mondal_fuller} had predicted the existence of large solvent fluctuations. Our MD thus qualitatively and quantitatively validates the prediction of that and previous works. In Table 1, we provide a summary of the timescales for all the cases obtained through metadynamics, unbiased MD and through the use of detailed balance.

\subsection{Ligand free to move in any direction}
\subsubsection{Thermodynamic profile}
Fig. 4(a) shows the 2-d free energy as a function of $z$ and $\rho = \sqrt{x^2 + y^2}$ obtained through metadynamics. For all practical purposes this free energy surface is indistinguishable from the one obtained through umbrella sampling and reported in SI. In SI we also provide a comparison of the 1-d free energy as a function of $z$ obtained from metadynamics and from umbrella sampling. Note from Fig. 4 that when constraints are lifted, the free energy minimum is now slightly off-center and deeper than the free energy at the geometric center of the cavity.  We find that for this case the system avoids the central dewetting pathway and instead takes an alternate route, rolling in and out along the sides of the cavity.  This pathway is favorable when compared with the axial symmetric path as the number of hydrophobic contacts between the ligand and the cavity are optimized.  Indeed, even though the minimum is now deeper than the sterically constrained case, the overall barrier is smaller (Fig. 4 and Fig. 1(b) in SI), and as we show in next subsection Kinetics, the unbinding is faster.

\begin{figure}[h!]
\label{fes_free}
  \centering
        \includegraphics[height=2in]{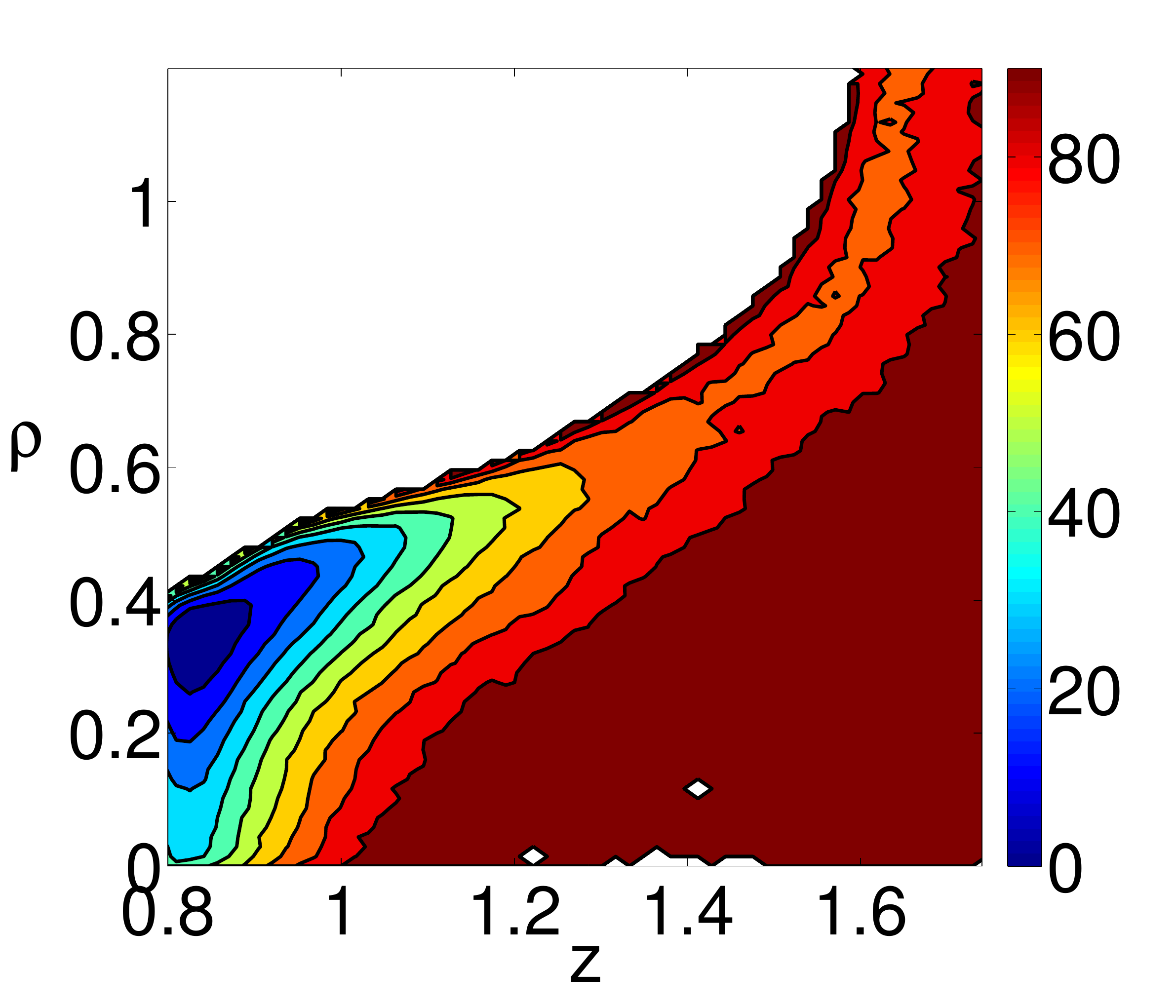}
        \includegraphics[height=2in]{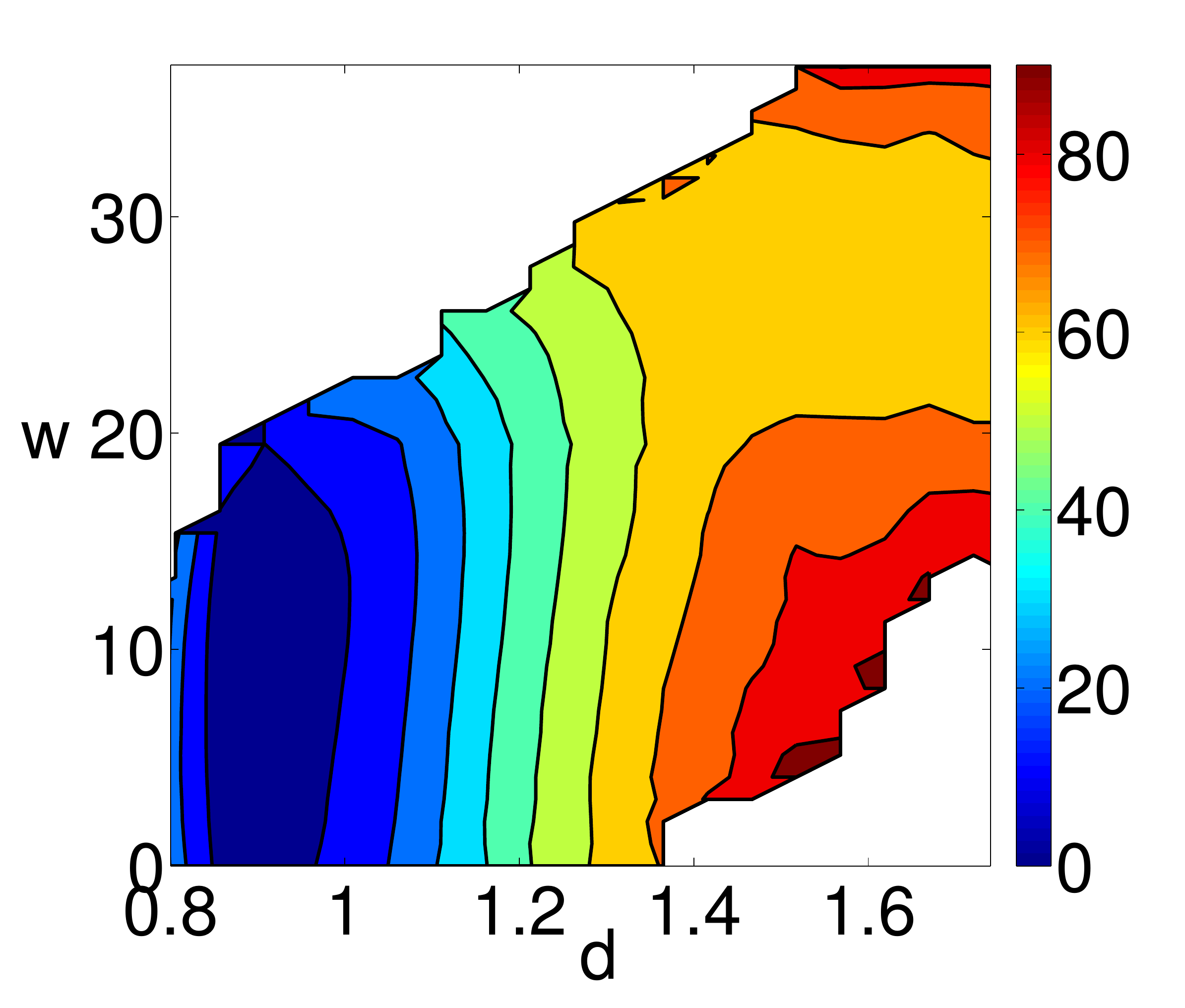}
        \captionsetup{justification=raggedright,
singlelinecheck=false
}
\caption{Various free energy profiles from metadynamics for the case when the ligand is free to move in any direction. (a) is  as function of ($z$,$\rho$), while (b) is as a function of ($d$,$w$). Note the clear lack of a sharp dewetting transition in (b), in contrast with Figs. 2 and 3. All energies are in kJ/mol and contours are separated by 10 kJ/mol.}
\end{figure}

Fig. 4(b) shows the 2-d free energy but now as a function of ligand-cavity separation $d=\sqrt {x^2 + y^2 + z^2}$  and the pocket water occupancy $w$. This is to be contrasted with Fig. 2 for the sterically constrained motion case. This clearly demonstrates that the water occupancy in this case changes gradually without any sharp transitions at specific cavity-ligand separations. 

\subsubsection{Kinetics}
To obtain unbinding kinetics for this case, we perform 14 independent infrequent 1-d metadynamics biasing $d=\sqrt {x^2 + y^2 + z^2}$, starting from fully bound ligand and stopping when unbound (see Methods and SI for detailed simulation parameters). For the current purpose, we take the unbound ligand as $z=1.5$, $\rho = 0.6-0.8$. The unbinding time now reduces to 200 $\pm$ 51 sec, with again an excellent Poisson fit demonstrating reliability of the choice of CV and the timescales so generated. Using the free energy difference of -69.8 kJ (Fig 1(b) in SI) between the bound and unbound states and the principle of detailed balance, we obtain a mean binding time of 118 $\pm$ 31 ps. This is again in good agreement with the estimate of 157 $\pm$ 5 ps through unbiased MD.

%------------------------------------------------

\section{\large{Discussion}}
In this paper, we have provided detailed thermodynamic and kinetic insights into the binding and unbinding of a popular prototypical ligand-substrate system through fully atomistic metadynamics-assisted molecular dynamics simulations performed in explicit water. Such systems are very relevant to the understanding of a range of processes across biology, chemistry, and biochemistry. We find that in accordance with previous Brownian dynamics based studies \cite{mondal_fuller,morrone2012interplay,li2012hydrodynamic}, the binding of such a system proceeds through a marked dewetting transition, but only when the system is sterically constrained to move along its axis of symmetry. We calculated the unbinding time to be around 4000 secs, with an associated barrier of roughly 80 kJ/mol. When the steric constraint is removed, the dewetting transition becomes continuous, and the unbinding proceeds through an alternate preferred pathway in which the residence time of the ligand decreases 20-fold to 200 secs. These extremely long unbinding timescales were obtained through the use of metadynamics with its recent extension\cite{meta_time}, and were validated through alternate simulation techniques and the principle of detailed balance. 

Even though the systems considered here are far simpler than an actual complex drug-protein system, we believe this work has multiple useful implications. Firstly, this is one of the first times that such a quantitatively insightful study has been carried on realistic ligand-substrate systems with very slow unbinding kinetics, and validated through a range of simulation techniques. The relative simplicity of the system allowed us to validate the timescales through detailed balance by performing thorough unbiased MD simulations of the binding, thereby giving confidence in the use of metadynamics type technique for getting unbinding kinetics in more complex protein-ligand systems\cite{trypsin}. Secondly, our work shows how the presence of simple steric constraints can heavily influence the role played by molecular solvent. We hope that our work will serve as a useful addition to the understanding of hydrophobic interactions in solvated systems in biology and chemistry.

\begin{acknowledgments}
 This work was supported by grants from the National Institutes of Health [NIH-GM4330] and the Extreme Science and Engineering Discovery Environment (XSEDE) [TG-MCA08X002].
\end{acknowledgments}

\bibliographystyle{pnas}
\bibliography{fullerene_arxiv}

\begin{thebibliography}{10}

\bibitem{copeland}
Copeland RA, Pompliano DL, Meek TD (2006) Drug--target residence time and its
  implications for lead optimization.
\newblock {\em Nat Rev Drug Discov} 5:730--739.

\bibitem{pan_kinetics}
Pan AC, Borhani DW, Dror RO, Shaw DE (2013) Molecular determinants of
  drug--receptor binding kinetics.
\newblock {\em Drug Discov Today} 18(13):667--673.

\bibitem{mondal_fuller}
Mondal J, Morrone JA, Berne BJ (2013) How hydrophobic drying forces impact the
  kinetics of molecular recognition.
\newblock {\em Proceedings of the National Academy of Sciences} 110:13277.

\bibitem{setny2013solvent}
Setny P, Baron R, Kekenes-Huskey PM, McCammon JA, Dzubiella J (2013) Solvent
  fluctuations in hydrophobic cavity--ligand binding kinetics.
\newblock {\em Proceedings of the National Academy of Sciences}
  110(4):1197--1202.

\bibitem{debenedetti_2015}
Remsing RC et~al. (2015) Pathways to dewetting in hydrophobic confinement.
\newblock {\em Proceedings of the National Academy of Sciences}
  112(27):8181--8186.

\bibitem{morrone2012interplay}
Morrone JA, Li J, Berne BJ (2012) Interplay between hydrodynamics and the free
  energy surface in the assembly of nanoscale hydrophobes.
\newblock {\em The Journal of Physical Chemistry B} 116:378.

\bibitem{li2012hydrodynamic}
Li J, Morrone JA, Berne B (2012) Are hydrodynamic interactions important in the
  kinetics of hydrophobic collapse?
\newblock {\em The Journal of Physical Chemistry B} 116(37):11537--11544.

\bibitem{sharma2012evaporation}
Sharma S, Debenedetti PG (2012) Evaporation rate of water in hydrophobic
  confinement.
\newblock {\em Proceedings of the National Academy of Sciences}
  109(12):4365--4370.

\bibitem{giovambattista2006effect}
Giovambattista N, Rossky PJ, Debenedetti PG (2006) Effect of pressure on the
  phase behavior and structure of water confined between nanoscale hydrophobic
  and hydrophilic plates.
\newblock {\em Physical Review E} 73(4):041604.

\bibitem{mccammon_jacs}
Baron R, Setny P, Andrew~McCammon J (2010) Water in cavity−ligand
  recognition.
\newblock {\em Journal of the American Chemical Society} 132(34):12091--12097.
\newblock PMID: 20695475.

\bibitem{young2007motifs}
Young T, Abel R, Kim B, Berne BJ, Friesner RA (2007) Motifs for molecular
  recognition exploiting hydrophobic enclosure in protein--ligand binding.
\newblock {\em Proceedings of the National Academy of Sciences}
  104(3):808--813.

\bibitem{trypsin}
Tiwary P, Limongelli V, Salvalaglio M, Parrinello M (2015) Kinetics of
  protein–ligand unbinding: Predicting pathways, rates, and rate-limiting
  steps.
\newblock {\em Proceedings of the National Academy of Sciences}
  112(5):E386--E391.

\bibitem{hummer_review}
Rasaiah JC, Garde S, Hummer G (2008) Water in nonpolar confinement: From
  nanotubes to proteins and beyond.
\newblock {\em Annu. Rev. Phys. Chem.} 59:713--740.

\bibitem{hiv}
Braaten D, Ansari H, Luban J (1997) The hydrophobic pocket of cyclophilin is
  the binding site for the human immunodeficiency virus type 1 gag polyprotein.
\newblock {\em Journal of virology} 71(3):2107--2113.

\bibitem{shaw_dasatanib}
Shan Y et~al. (2011) How does a drug molecule find its target binding site?
\newblock {\em Journal of the American Chemical Society} 133(24):9181--9183.
\newblock PMID: 21545110.

\bibitem{mondal_dasatanib}
Mondal J, Friesner RA, Berne BJ (2014) Role of desolvation in thermodynamics
  and kinetics of ligand binding to a kinase.
\newblock {\em Journal of chemical theory and computation} 10(12):5696--5705.

\bibitem{ladbury1996}
Ladbury JE (1996) Just add water! the effect of water on the specificity of
  protein-ligand binding sites and its potential application to drug design.
\newblock {\em Chemistry \& biology} 3(12):973--980.

\bibitem{tuckerman1992}
Tuckerman M, Berne BJ, Martyna GJ (1992) Reversible multiple time scale
  molecular dynamics.
\newblock {\em The Journal of chemical physics} 97(3):1990--2001.

\bibitem{willard_tps}
Willard AP, Chandler D (2008) The role of solvent fluctuations in hydrophobic
  assembly.
\newblock {\em The Journal of Physical Chemistry B} 112(19):6187--6192.

\bibitem{bolhuis_tps}
Bolhuis PG, Chandler D (2000) Transition path sampling of cavitation between
  molecular scale solvophobic surfaces.
\newblock {\em The Journal of Chemical Physics} 113(18):8154--8160.

\bibitem{gardehummerprl}
Hummer G, Garde S (1998) Cavity expulsion and weak dewetting of hydrophobic
  solutes in water.
\newblock {\em Physical review letters} 80(19):4193.

\bibitem{meta_laio}
Laio A, Parrinello M (2002) Escaping free-energy minima.
\newblock {\em Proc Natl Acad Sci} 99(20):12562--12566.

\bibitem{wtm}
Barducci A, Bussi G, Parrinello M (2008) Well-tempered metadynamics: A smoothly
  converging and tunable free-energy method.
\newblock {\em Phys Rev Lett} 100(2):020603--020606.

\bibitem{tiwary_rewt}
Tiwary P, Parrinello M (2015) A time-independent free energy estimator for
  metadynamics.
\newblock {\em J Phys Chem B}.

\bibitem{meta_ode}
Dama JF, Parrinello M, Voth GA (2014) Well-tempered metadynamics converges
  asymptotically.
\newblock {\em Phys Rev Lett} 112(24):240602--240605.

\bibitem{meta_time}
Tiwary P, Parrinello M (2013) From metadynamics to dynamics.
\newblock {\em Physical review letters} 111(23):230602.

\bibitem{pvalue}
Salvalaglio M, Tiwary P, Parrinello M (2014) Assessing the reliability of the
  dynamics reconstructed from metadynamics.
\newblock {\em J Chem Theory Comp} 10(4):1420--1425.

\bibitem{berneweekszhou}
Berne BJ, Weeks JD, Zhou R (2009) Dewetting and hydrophobic interaction in
  physical and biological systems.
\newblock {\em Annual review of physical chemistry} 60:85.

\bibitem{plumed1}
Bonomi M et~al. (2009) Plumed: A portable plugin for free-energy calculations
  with molecular dynamics.
\newblock {\em Comp Phys Comm} 180(10):1961--1972.

\bibitem{meta_review}
Barducci A, Bonomi M, Parrinello M (2011) Metadynamics.
\newblock {\em Wiley Interdisciplinary Reviews: Computational Molecular
  Science} 1(5):826--843.

\bibitem{bonomi_rewt}
Bonomi M, Barducci A, Parrinello M (2009) Reconstructing the equilibrium
  boltzmann distribution from well-tempered metadynamics.
\newblock {\em J Comp Chem} 30(11):1615--1621.

\bibitem{besthummer_rc}
Best RB, Hummer G (2005) Reaction coordinates and rates from transition paths.
\newblock {\em Proceedings of the National Academy of Sciences of the United
  States of America} 102(19):6732--6737.

\bibitem{berne1988}
Berne BJ, Borkovec M, Straub JE (1988) Classical and modern methods in reaction
  rate theory.
\newblock {\em J. Phys. Chem.} 92(13):3711--3725.

\bibitem{voter_prl}
Voter AF (1997) Hyperdynamics: Accelerated molecular dynamics of infrequent
  events.
\newblock {\em Phys Rev Lett} 78:3908--3911.

\bibitem{grub_pre}
Grubm\"uller H (1995) Predicting slow structural transitions in macromolecular
  systems: Conformational flooding.
\newblock {\em Phys Rev E} 52:2893--2906.

\bibitem{throwingropes}
Bolhuis PG, Chandler D, Dellago C, Geissler PL (2002) Transition path sampling:
  Throwing ropes over rough mountain passes, in the dark.
\newblock {\em Ann Rev Phys Chem} 53(1):291--318.

\end{thebibliography}

\end{document}